\documentclass[12pt]{iopart}
\begin{document}
\title[Fermi and identical particles]{Enrico Fermi's view of identical particles.}

\author{Edoardo Milotti}

\address{Dipartimento di Fisica dell'Universit\`a di Trieste, and I.N.F.N. -- Sezione di Trieste, Via Valerio, 2 -- I-34127 Trieste, Italy}
\ead{milotti@ts.infn.it}

\begin{abstract}
In this paper I discuss Enrico Fermi's view of identical particles, taking  a lecture that he gave in 1933 as a starting point.  Fermi used his lecture as a basis for a paper that was published in 1934: the paper is in italian and is not easily accessible to a wide audience, and for this reason its translation is also given in a section of the present paper.
\end{abstract}
\maketitle

\section{Introduction}

Enrico Fermi has long thought about identical particles in his effort to understand the behavior of entropy in quantum systems. In the paper "Considerations on the quantization of systems that contain identical elements", published in italian in 1924 \cite{ef1}, he actually came very close to an independent formulation of the exclusion principle, which was put forward one year later by Pauli \cite{wp1}. For this reason he was well-prepared when Pauli published his paper \cite{wp1}, and then he developed quickly and independently the statistics that bears his name \cite{ef2}.  During the year 1933, Enrico Fermi gave a lecture in Bari, at the 22nd meeting of the {\it Societ\`a Italiana per il Progresso delle Scienze} ({\it Italian Society for the Progress of the Sciences}) \cite{sips}, and later published it in the journal {\it Scientia} under the title ``Le ultime particelle costitutive della materia'' (``The ultimate constituents of matter'') \cite{ef}. In his paper he described in his usual way, both thorough and simple, the particle world, as it was understood at that time, and he also made many interesting comments on identical particles and the exclusion principle.

In the next two sections I comment on those parts of Fermi's paper \cite{ef} that are relevant to the discussion on identical particles, while section \ref{trans} is the complete translation into english of the paper, which is in italian and is not easily accessible.

\section{Identical particles}

In 1933 the exclusion principle for electrons was well established and accepted. Pauli had formulated it in 1925 \cite{wp1}, and already in 1926 Heisenberg had derived the basic consequences of the principle on the wavefunction of a system of particles of the same kind \cite{wh}. The exclusion principle is obviously related to the existence of identical particles, and in his lecture Fermi remarks that this identity is something that belongs to the world of the basic constituents of matter. He argues that in our ordinary world there cannot be identical objects, because all macroscopic objects contain many different parts, and thus have an exceedingly large number of different states, so large in fact that no two similar objects can actually be found in the same state, and thus have different properties (e.g. mass, which depends on the interaction energy) even though they have the same parts. 
The Pauli exclusion principle is so basic to our understanding of the physical world, and its standing as a global symmetry is so unique among the physical laws, that its absolute validity is still debated \cite{ds}. The idea that electrons may be composite systems that contain a fermion and a boson has recently been explored by Akama et al. \cite{akama}, and it leads to the interesting conclusion that this kind of compositeness shows up as an apparent violation of both the Pauli exclusion principle -- in the same perspective put forward by Fermi -- and to an apparent violation of microcausality (and this is related to some specific features of string theory \cite{og} and also of any theory that incorporates the holographic principle \cite{busso}).

In the lecture Fermi also suggests that electrons may be a little bit different, but that the difference could be so tiny that it would be impossible to assess with actual measurements. This is exactly the view that was advocated much later in theories of violation of the Pauli exclusion principle \cite{ik,bieden}, however this is tenable only in the approximation of a non-relativistic theory, and as soon as the creation and annihilation of particles is also included (i.e., in a quantum field theory) this remark is no longer true. This was noted in 1989 by Govorkov \cite{gov}, who proved that a three-level Fermi oscillator like that of Ignatiev and Kuzmin \cite{ik} leads to states with negative norm. Another poignant remark was made by Amado and Primakoff \cite{ap}, who noticed that even a single slightly different kind of electron would lead to a doubling of the high-energy e$^+$- e$^-$ production cross-section in e$^+$- e$^-$ colliders. Thus, the idea of a small difference suggested by Fermi seems to be wrong, but in a very subtle way.

\section{Further remarks}

The exclusion principle was introduced by Pauli to explain the observed lengths of the periods of the atomic system of the elements \cite{wpn}, and therefore the electron was taken as a fermion from the very beginning. 
However the statistics of nuclei was not as obvious. The problem could be solved with the investigation of band spectra in molecules with two like atoms, since these spectra display lines with alternating intensity, and depend on the symmetry of the nuclear wavefunction. In this way it was possible to establish that the proton follows the Fermi-Dirac statistics. It was more difficult to pinpoint the identity of the other atomic nuclei -- which were believed to be bound states of protons and electrons -- since most cases seemed well-understood, while the N$_2$ molecule behaved in the opposite way (intense lines should have been weak and vice versa -- the situation is well summarized by Ehrenfest and Oppenheimer in their 1931 paper \cite{ep}). In his paper \cite{ef} Fermi mentions the beautiful measurements on molecular nitrogen that were done by Rasetti in 1929 while he was at Caltech \cite{rasetti1,rasetti2}, and which established the ``nitrogen paradox'' on a firm experimental ground. 

As Fermi remarks in the paper, the nitrogen paradox was solved with the discovery of the neutron in 1932 \cite{chad}. It is interesting to notice that in 1933 Rasetti rejoined the Fermi group in Rome after one year at Caltech and a stay in Berlin (with Lise Meitner and Otto Hahn) \cite{rasetti1}. Rasetti was an excellent experimentalist, and after applying spectroscopic techniques to the newly-discovered Raman effect in Caltech, he went on to learn how to produce efficient radioactive sources under the caring supervision of Meitner and Hahn. Thus at the time of the 1933 lecture Fermi and his colleagues had all the tools -- theoretical and experimental -- that they needed to start their measurements with neutrons, a path that eventually led Fermi to the construction of the first nuclear reactor.

Even though the discovery of the neutron swept away the idea that nuclei contain proton-electron bound states and at the same time fixed the statistics of nuclei, nuclear $\beta$-emission was still unexplained, and seemed to indicate that electrons were somehow present in nuclei. At the end of 1933 Fermi reconciled these experimental facts with his theory of neutron $\beta$-decay. He was fully conscious of the importance of the theory \cite{segre}, but little of this transpires from the paper \cite{ef}.

Fermi devotes the concluding paragraphs of  his paper to the positron: Anderson discovered the positron and published his findings at the end of february 1933 \cite{and}, and Fermi explains very clearly how this discovery gives credibility to Dirac's hypothesis of the electron sea. This is directly related to the exclusion principle and is an essential component of Pauli's proof of the spin-statistics theorem \cite{spinstat}. Even though it is impossible to surmise that Fermi understood at this early stage the importance of this detail in the proof of the spin-statistics theorem, it is certainly intriguing to find that most of the essential physical aspects of the spin-statistics connection are so neatly summarized in a popular science paper.

\section{Translation of the paper by E. Fermi}
\label{trans}

The text that follows is the translation of the original paper \cite{ef} into English.
\bigskip

{\small
{\bf The ultimate constituents of matter}
\smallskip

The most essential difference between the objects of the macroscopic world, i.e. the common objects, and the microscopic objects might be the following: 
In the macroscopic world there are no two identical objects. Take for instance two iron blocks: we shall never be able to make them the same weight and the same shape; we might also try to forge them with the same microcrystalline structure, with the same annealing, with the same impurity content and so on. It is clear however that we shall never be able to make them quite the same, and the reason of this impossibility stems from the extreme complexity of these objects, which are clusters of billions of billions of atoms and molecules: a simple shift of one atom in one of the two iron blocks and they can no longer be identical. In this sense the non-existence of identical objects in the macroscopic world hints at a very complex structure.
The situation becomes quite different if we move from common objects to atoms and molecules, and even more so if we consider their constituents, nuclei and electrons. In fact we often meet identical objects in the atomic world; thus we can say for instance that any two electrons, or even any two atoms of the same kind, are identical. One might naturally object that the absolute identity between two objects is impossible to assess: it can be tested only in an approximate range that depends on our ability of observation. And it is unthinkable that this range, even when our technical means progressively improve, eventually vanishes. All this is obviously true and thus even the statement that any two electrons are the same must be relative to our present abilities of observation. And yet these abilities, in the specific case of the identity of two electrons, are extremely accurate. Skipping all details that would take us much too far, I shall only notice that the great precision with which we can assess the identity of two electrons depends on the possibility of observing the cumulative effects of an eventual difference in a long time span. For instance, if there were a difference, even a very slight one, between two electrons in the same atom, its effect would also be very small during a short period of the atom's life, but it would visibly change the structure and the external properties of the atom in the long term. Therefore we can state the identity of two electrons, if not in an absolute sense, at least in an extraordinarily small range.
Identical objects are often met in the atomic world and this encourages us to think that the structure of the atomic corpuscles is not extremely complex, and that as soon as the nature of the atomic corpuscles, nuclei and electrons, is clarified, we shall find that we have not merely shifted the problem of the structure of matter towards more minute components, but that we have reached a kind of plateau that might not yet represent the ultimate basis on which the scaffold of matter is built, a basis that might never be grasped by the human intelligence, but that could still be sufficient for a very long time. Naturally all this is just a conjecture, even though there are serious reasons to deem it a likely one.
Today the structure of the atom is fairly well known and we know that every atom contains a nucleus and a variable number of electrons and we can count the number of different particles of which all bodies are made up. Overall there are about a hundred different corpuscles, and this, as we noted above, seems to hint that we are not far from a complete analysis of all of the structural elements of matter. The electrons, which are present in all atoms, may be the most important ones; they are also those that have been known for the longest time, as they have been produced and studied in the electrical discharges in rarefied gases; the so-called cathode rays are indeed just a projection of very fast free electrons and, as everybody knows, from the study of their properties we can determine the charge and mass of the electron.
It is generally held that the electron is a true elementary corpuscle, i.e. that it cannot be further analyzed into simpler elements; and certainly up to now there is no indication of a more complex structure. On the other hand there are good reasons to suppose that the positive atomic nuclei are not simple corpuscles but clusters, in some cases quite crowded, of simpler elements.
An indication of this comes from the relatively large number of possible nuclei, which points to a somewhat complicated structure. But nuclear disintegrations, both those that occur spontaneously in radioactive substances, and the artificial ones that are produced bombarding a nucleus with alpha particles or very fast protons, show much more directly the complexity of nuclei. In all these cases we witness a breakdown or a rearrangement of the nuclear array; a nucleus can either emit or absorb corpuscles thereby changing into a nucleus of a different species. Today the structure of the atom is basically known, and the problem of analyzing and understanding the structure of nuclei has become the most central and deepest of physics studies.
Certainly, the simplest of all nuclei is the hydrogen nucleus or proton; it is the nucleus with the smallest mass (about one unit of atomic weight) and the smallest charge (equal to the electron's charge, but with opposite sign). The proton, just like the electron, is generally believed to be elementary, not made up of smaller parts; in any case it is for sure one of the fundamental elements of the nuclear structure.
Until two years ago, the electron and the proton were the only known simple corpuscles, and nuclei were believed to be just clusters of electrons and protons in such number as to give the right values of charge and mass of the nucleus. Today the perspectives for a theory of nuclei are vastly larger after the discovery of two new elementary, or at least very simple, particles: the neutron and the positive electron or positron. 
The neutron was the first to be discovered. First Bothe, in 1931, observed that Beryllium, when bombarded with alpha particles, emitted a radiation that was far more penetrating than all the known gamma ray radiations known at that time; shortly afterwards F. Joliot and his wife Irene Curie, daughter of the renowned discoverer of Radius, observed that the radiations emitted by Beryllium in the stated conditions had the property of scattering protons out of paraffin and other substances that contained hydrogen. This property showed that the radiation from Beryllium included something else than just gamma rays, which could transfer a very small momentum to relatively heavy corpuscles like protons. At the beginning of 1932 Chadwick, starting where Joliot had left, was able to show that the radiation from Beryllium could transfer momentum even to nuclei heavier than protons and interpreted these results, reaching the conclusion that the radiation involved a new kind of particles, which had often been suspected to exist but had never been found before, and that he called neutrons. They are electrically neutral, and their mass is very close to the proton mass, i.e. about one unit of atomic weight. The strong penetrating power of neutrons, which can go through several centimeters of lead, is easily explained from their being electrically neutral; for this reason the electric fields of the electrons in a body, which would otherwise have affected them, are ineffective, and neutrons can only react with a nucleus, when they pass extremely close by. Indeed, the production of a nuclear disintegration by throwing a positively charged corpuscle, such as a proton or an alpha particle, against the nucleus is hampered by the positive charge of the nucleus which repels the incoming particle; while a neutron, which is uncharged, is not repelled and can easily reach the nucleus.
The discovery of the other elementary particle which we mentioned above, the positron or positive electron, was announced for the first time by Anderson about one year ago; a few months later the discovery was confirmed and demonstrated in a much more complete way by Blackett and Occhialini. These physicists, observing the disintegrations of matter produced by corpuscles of the cosmic radiation, witnessed sometimes real explosions in which a nucleus blew up in some twenty corpuscles projected outwards like bullets in a shrapnel. By deflecting these corpuscles with a magnetic field they noticed that while some of them were bent like negative electrons, the trajectories of some other particles were bent in the opposite way, which meant that they were positively charged corpuscles. From the ionization density produced by these corpuscles, and from other considerations, it was shown that their mass had to be of the order of magnitude of the electron mass. This was indeed a positive electron. After the discovery of the positron, its presence was detected also in phenomena that did not depend on penetrating radiation. One particularly important discovery was that by irradiating atoms, especially the heavy ones, with very hard gamma radiations, positive electrons are produced, or rather, in most cases, a pair of one positive and one normal negative electron. We shall soon see the theoretical relevance of this fact.
The discovery of the neutron allows, according to Heisenberg and E. Majorana, to construct a general scheme of the structure of nuclei, free of many of those objections that could be raised against the primitive model in which nuclei were made up of protons and electrons only. Indeed, given the very small mass of these last corpuscles, one electron confined within an orbit with the size of a nucleus, should acquire huge kinetic energies that should be well visible in the atomic mass defect; in addition to this quantitative difficulty there are other qualitative arguments that is not possible to state here because of lack of space. I shall only mention that the statistical behavior of the nitrogen nucleus, first evidenced in the beautiful observations of Raman effect in gaseous nitrogen by Rasetti, seems to be incompatible with an odd number of elementary particles (14 protons and 7 electrons according to the old scheme); such incompatibilities, first observed in nitrogen, have later been found in many other examples, and have completely discredited the scheme with protons and electrons. 
According to the theories of Heisenberg and Majorana the fundamental elements of the nuclear structure should rather be protons and neutrons. The latter might in turn be the union of a proton with an electron, which might be necessary to explain the possibility of a disintegration with the emission of $\beta$-rays, but this would be an assembly made according to laws unlike those of ordinary quantum mechanics: just like ordinary mechanics loses its validity in the description of the behavior of the electrons when we move from the ordinary to the atomic scale, so the new scale change from atomic to nuclear phenomena would necessitate a new, as yet unknown, change; it seems however that ordinary quantum mechanics suffices to describe, even inside the nucleus, the behavior of relatively heavy corpuscles like protons and neutrons. Indeed they have a large mass and have, even in nuclear orbits, speeds that are quite small when compared with the speed of light, so that the relativistic corrections to their motion always remain of secondary importance. Therefore, if we shall indeed be able to show that the nuclear electrons, whose existence must be admitted to explain the $\beta$-particle emission, do not exist in the free state, but are for instance intimately associated to protons so as to form neutrons, a relatively easy possibility of constructing a fairly advanced theory of the nucleus shall remain open, inasmuch as it shall be possible, at least until we want to study those phenomena in which the neutronic structure is involved, to use the procedure and the general interpretative scheme of quantum mechanics. Obviously, as soon as we reach this point, even when we discount the mathematical difficulties, we would still have the essential problem of knowing the laws of the forces acting among the corpuscles in the nucleus. Later on we would still have to clarify the structure of the neutron, for which, as we said, quantum mechanics could probably no longer be applied; for this purpose we have before us a few clues, from the continuous $\beta$-ray spectrum, which according to Bohr might lead us to think that for these new unknown laws even the principle of conservation of energy might no longer hold; unless we follow Pauli and admit the existence of the so called ÇneutrinoÈ, i.e. a hypothetical particle, electrically neutral and with a mass of the order of magnitude of the electron mass. It would have an enormous penetrating power and would thus escape all present detection methods, and its kinetic energy would be used to reestablish the energy balance in $\beta$ disintegrations. Certainly the accumulation of knowledge about the various ways of disintegration of nuclei shall shine much light on all these questions; and hope is justified by the unforeseen speed of experimental progress in the last two years.
We must still add a few words on the theoretical importance of the discovery of the positive electron. All attempts to construct a quantum mechanics for the electron that were compatible with the relativity principle, attempts that were crowned by the work of Dirac, had always met the difficulty of explaining the appearance, along with normal states that describe the behavior of the electron, of other states, without an apparent physical match, in which the electron had a negative kinetic energy. Dirac proposed then to admit that all these anomalous states were occupied; space would then be uniformly filled, and because of its uniformity this filling would be unobservable; a hole in this uniform distribution could instead be observed, that is the lack of an electron in a negative energy state. Dirac showed that one of these holes would behave just like a normal corpuscle, with a positive electric charge, and originally put forward the hypothesis that the protons could be identified with these holes. Dirac's original hypothesis was shown to be untenable, because it was unable to explain the large mass difference between protons and electrons, and since at that time experience had not yet shown the existence of a corpuscle with the properties of Dirac's holes, it was believed that they had no real counterpart. The discovery of the positive electron has substantially modified this situation: the new corpuscle has in fact, at least as far as we know today, exactly those properties that Dirac's theory assigns to holes; it is natural then to identify holes with the positive electron. If this view is correct, it must be possible, extracting an electron from a negative energy state and bringing it to a positive energy state, to form simultaneously a hole, i.e. a positive electron and a normal negative electron. It seems that a phenomenon such as this actually happens in the formation of electron-positron pairs produced by hard gamma-rays. If it shall be shown that it is actually so, the positive and negative electrons that appear in this way may with good right be called, according to the proposal of Madame Curie, materialization electrons, as they represent the complete transformation of radiant energy into material corpuscles. }

\section{Final comments}

Fermi had a special gift of bringing light to the most obscure problems, and this shows up once again in this lecture which discusses problems that were at the forefront of research at that time. 

Emilio Zavattini, who as a young physicist attended the lectures that Fermi gave at the summer school in Varenna in 1954, during Fermi's last visit to Italy, witnesses that ``Fermi was a very kind person, who listened carefully to your problem, and had the uncanny ability of breaking it down to small pieces and then reassembling it into a new, simpler, and thoroughly clear shape. Even though he didn't mean it, this made you feel like a fool'' \cite{mimmo}.  Fermi's clarity of thought has led to firm concepts that stand very well the wear of time: even now, more than seventy years after paper \cite{ef}, Fermi's remarks on particle identity and on the exclusion principle are not outdated, and may point the way to an interpretation of modern tests of the exclusion principle (like, e.g., \cite{vip}).

\end{document}